\documentclass[twocolumn,prl,aps,superbib,tightenlines,floatfix,superscriptaddress,showpacs]{revtex4}
\usepackage{amsmath}
\usepackage{graphicx}
\usepackage{amssymb}
\usepackage{color,soul,ulem}
\usepackage[T1]{fontenc}
\usepackage{ae,aecompl}

\begin{document}

\title{Switching Synchronization in One-Dimensional Memristive Networks}

\author{Valeriy A. Slipko}
\affiliation{Department of Physics and Technology, V. N. Karazin Kharkov National University, Kharkov 61022, Ukraine}
\author{Mykola Shumovskyi}
\affiliation{Department of Physics and Technology, V. N. Karazin Kharkov National University, Kharkov 61022, Ukraine}
\author{Yuriy V. Pershin}
\email{pershin@physics.sc.edu}
\affiliation{Department of Physics and Astronomy and Smart State Center for Experimental Nanoscale Physics, University of South Carolina, Columbia, South Carolina 29208, USA}

\begin{abstract}
We report on an astonishing switching synchronization phenomenon in one-dimensional memristive networks, which occurs when several memristive systems with different switching constants are switched from the high to low resistance state. Our numerical simulations show that such a collective behavior is especially pronounced when  the applied voltage slightly exceeds the combined threshold voltage of memristive systems. Moreover, a finite increase in the network switching time is found compared to the average switching time of individual systems. An analytical model is presented to explain our observations. Using this model, we have derived asymptotic expressions for memory resistances at short and long times, which are in excellent agreement with our numerical calculations.\end{abstract}


\maketitle

{\it Introduction.}-- Synchronization is the term that is frequently used to describe the coherent dynamics of an ensemble of interconnected dynamical units, namely, dynamical units forming networks. The networks are ubiquitous in nature and technology, and, therefore, it is not surprising that the phenomenon of synchronization has been studied and observed in a wide range of dynamical systems. These systems include, for example, neurons \cite{Elson98a,Neiman02a}, power grids \cite{Rohden12a,Motter13a}, coupled lasers \cite{Nixon12a}, oscillators, social systems \cite{Neda00a}, etc. A lot of attention has be drawn to the synchronization of chaotic systems \cite{Pecora90a,Rosenblum96a} -- an intriguing emergence of collective dynamics of a number of chaotic units linked with a common signal or signals. Oscillator networks \cite{Dorfler13a} are another example of widely studied systems with synchronization.

In this Letter, our attention is focused on memristive (memory resistive) networks. These networks are composed of individual memristive elements \cite{chua76a}, which now are of considerable interest for a variety of applications. In these passive resistive electronic devices, the resistance depends on the history of signals applied. The ability of memristive systems to store and process information on the same physical platform makes them ideal for unconventional computing applications \cite{diventra13a,pershin12c}. In fact,
boolean logic operations with small memristive networks were experimentally demonstrated few years ago \cite{borghetti10a}. Moreover, it was theoretically shown that larger memristive networks could solve maze \cite{pershin11d} and shortest path optimization \cite{pershin13b} problems in a single step compared to multi-step algorithms employed in conventional computers. Therefore, it's of a real importance to understand the dynamical properties of memristive networks.

Recently, two of us (VAS and YVP) have found  that in one-dimensional memristive networks subjected to adiabatically increasing voltage,
the effective switching rates of memristive systems strongly depend on their polarities \cite{pershin13a}. It has been demonstrated (on the level of individual memristive elements) that an abrupt (accelerated) switching occurs when the memristance (memory resistance) of a given memristive system in the network increases at the given voltage polarity. A slow (decelerated) switching takes place in the opposite case \cite{pershin13a}. However, this prior work leaves open the question of the switching behavior beyond the adiabatic limit, namely, when the applied voltage is initially high enough to induce the dynamics of several memristive systems. This is precisely the aim of the present Letter, which explores the switching dynamics of one-dimensional memristive networks subjected to sufficiently high voltages. According to our findings, an interesting switching synchronization effect takes place when  all memristive systems switch from the high to low resistance state. Our consideration of the switching synchronization effect employs both numerical and analytical techniques.

\begin{figure}[b]
\begin{center}
\includegraphics[width=.75\columnwidth]{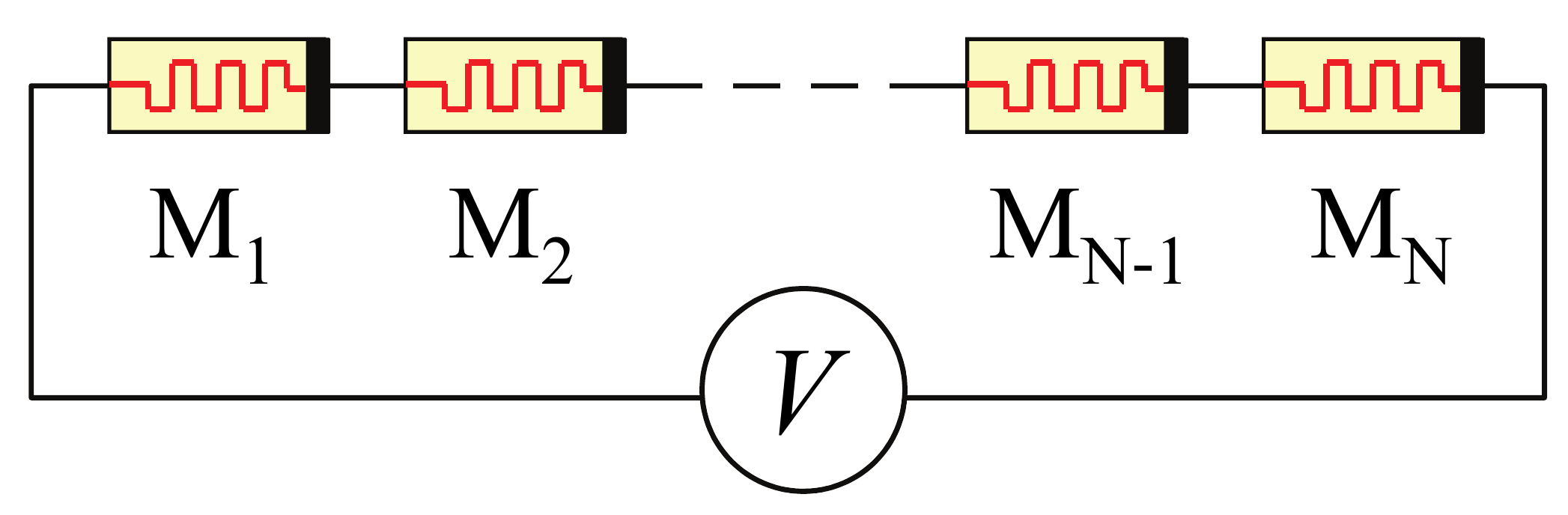}
\caption{\label{fig1} (Color online) One-dimensional network of $N$ memristive systems M$_i$ connected to a voltage source $V$. Here we assume that all the memristive systems M$_i$ are connected with the same polarity and $R_i(t=0)=R_{off}$, where $R_{off}$ is the high resistance state.}
\end{center}
\end{figure}

Fig. \ref{fig1} shows a one-dimensional memristive network connected to a constant voltage source $V$. It is assumed that the polarities of all memristive systems are the same, and, at the initial moment of time $t=0$, all memristive systems are in the same high resistance state $R_i(t=0)=R_{off}$. Such configuration is similar to the initial configurations in the maze \cite{pershin11d} and shortest path problem \cite{pershin13b} solving networks and thus of importance for potential applications of memristive systems.

We assume that Fig. \ref{fig1} network employs voltage-controlled memristive systems with threshold that have been experimentally realized with different materials combinations \cite{pershin11a}. In our numerical simulations and analytical calculations presented below, we use a  model of voltage-controlled memristive systems with threshold \cite{pershin13a,pershin09b}
\begin{eqnarray}
I&=&R_i^{-1}(x_i)V_i \label{eq:model1} \\
\frac{\textnormal{d}x_i}{\textnormal{d}t}&=& \begin{cases} \pm \textnormal{sign}(V_i)\beta_i(|V_i|-V_t) \;\textnormal{if} \;\; |V_i|>V_t \\ 0 \;\;\;\;\;\;\;\;\;\;\;\; \textnormal{otherwise} \end{cases} , \label{eq:model2}
\end{eqnarray}
where $I$ and $V_i$ are the current through and the voltage across $i$-th memristive system, respectively, $x_i$ is the internal state variable playing
the role of memristance, $R_i(x_i)\equiv x_i$, $\beta_i$ is a positive switching constant characterizing the intrinsic rate of memristance change when $|V_i|>V_t$, $V_t$ is the threshold voltage, and $+$ or $-$ sign is selected according to the device connection polarity. Additionally, it is assumed that the memristance is limited to the interval [$R_{on}$, $R_{off}$] (note that $R_{on}< R_{off}$).

\begin{figure}[tb]
\begin{center}
\includegraphics[width=.95\columnwidth]{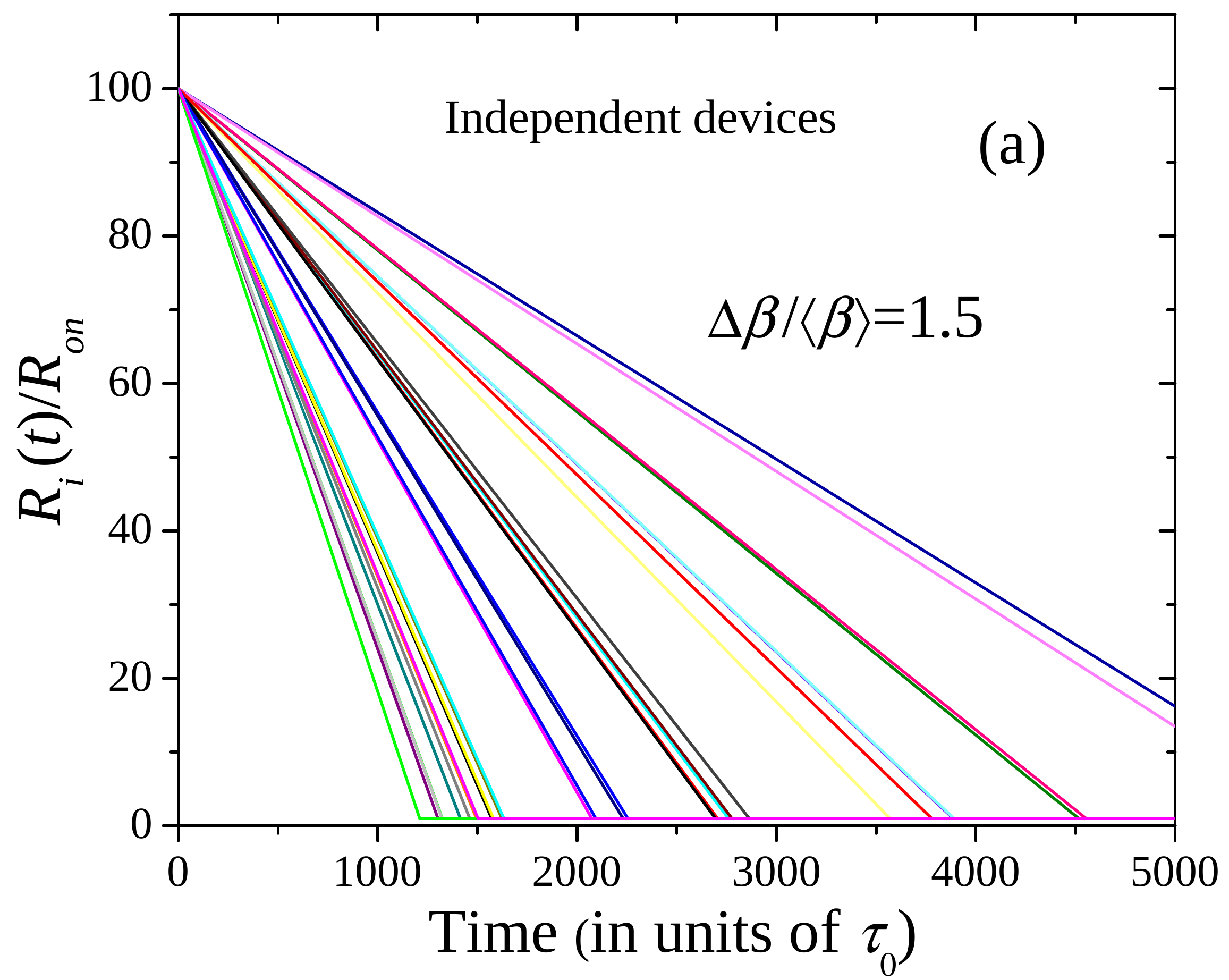}
\includegraphics[width=.95\columnwidth]{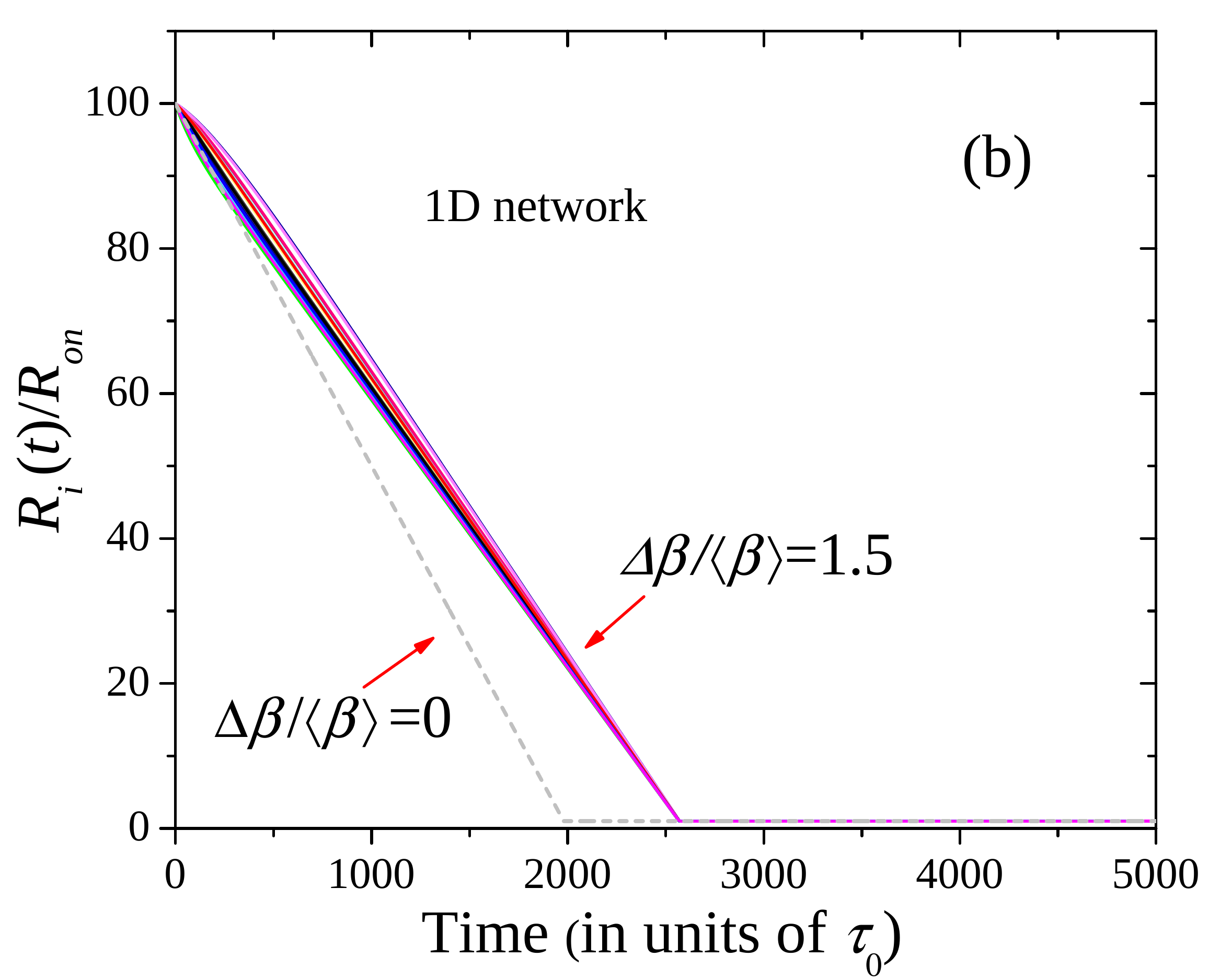}
\caption{\label{fig2} (Color online) The memristance $R_i(t)$ of $N=30$ memristive systems with $R_{off}/R_{on}=100$ (a) individually subjected to the same voltage $V_i=1.05V_t$, and (b) forming 1D network (as in Fig. \ref{fig1}) subjected to $V=1.05NV_t$. These plots have been obtained with a random flat distribution of parameters $\beta_i$ in the interval $[\langle\beta\rangle-\Delta \beta/2,\;\;\langle \beta\rangle+\Delta \beta/2]$. The time is measured in units of $\tau_0=R_{on}/(\langle\beta\rangle V_t).$}
\end{center}
\end{figure}

{\it Numerical Results.}-- Let us, first of all, consider switching in an ensemble of individual memristive systems characterized by a probabilistic distribution of the parameter $\beta_i$. For the sake of simplicity, we consider a flat random distribution of this parameter keeping all other parameters of memristive systems the same. Fig. \ref{fig2}(a) shows the time-dependence of memristances in an ensemble of memristive systems each subjected to the same voltage $V_i=1.05V_t$, $i=1,..,N$, which is the average voltage per system in the network considered in the next paragraph. It's not surpising that the switchings of these memristive systems occur at very different rates defined by specific individual values of switching constants $\beta_i$.


Next, we consider the collective switching, namely, the switching of memristive systems that form one-dimensional networks as the one sketched in Fig. \ref{fig1}. Fig. \ref{fig2}(b) shows the main result of this Letter. This plot demonstrates a surprising switching synchronization effect in which the effective switching rates of unlike systems become the same. This is a truly remarkable behavior that has not been anticipated in the literature. The basic principles of this behavior are related to the phenomenon of the decelerated switching \cite{pershin13a} that, unlike our previous investigation \cite{pershin13a}, takes place simultaneously in every component of the network. Technically speaking, the switching of memristive systems with larger values of  $\beta_i$  can not proceed fast as the decreases of their memristances also suppress the voltage falls across them. At the same time, the voltage falls across memristive systems with smaller values of $\beta_i$ increase compensating the smallness of their $\beta_i$. As a result, the switching of all memristive system occurs coherently with approximately the same effective rate.

\begin{figure}[tb]
\begin{center}
\includegraphics[width=.85\columnwidth]{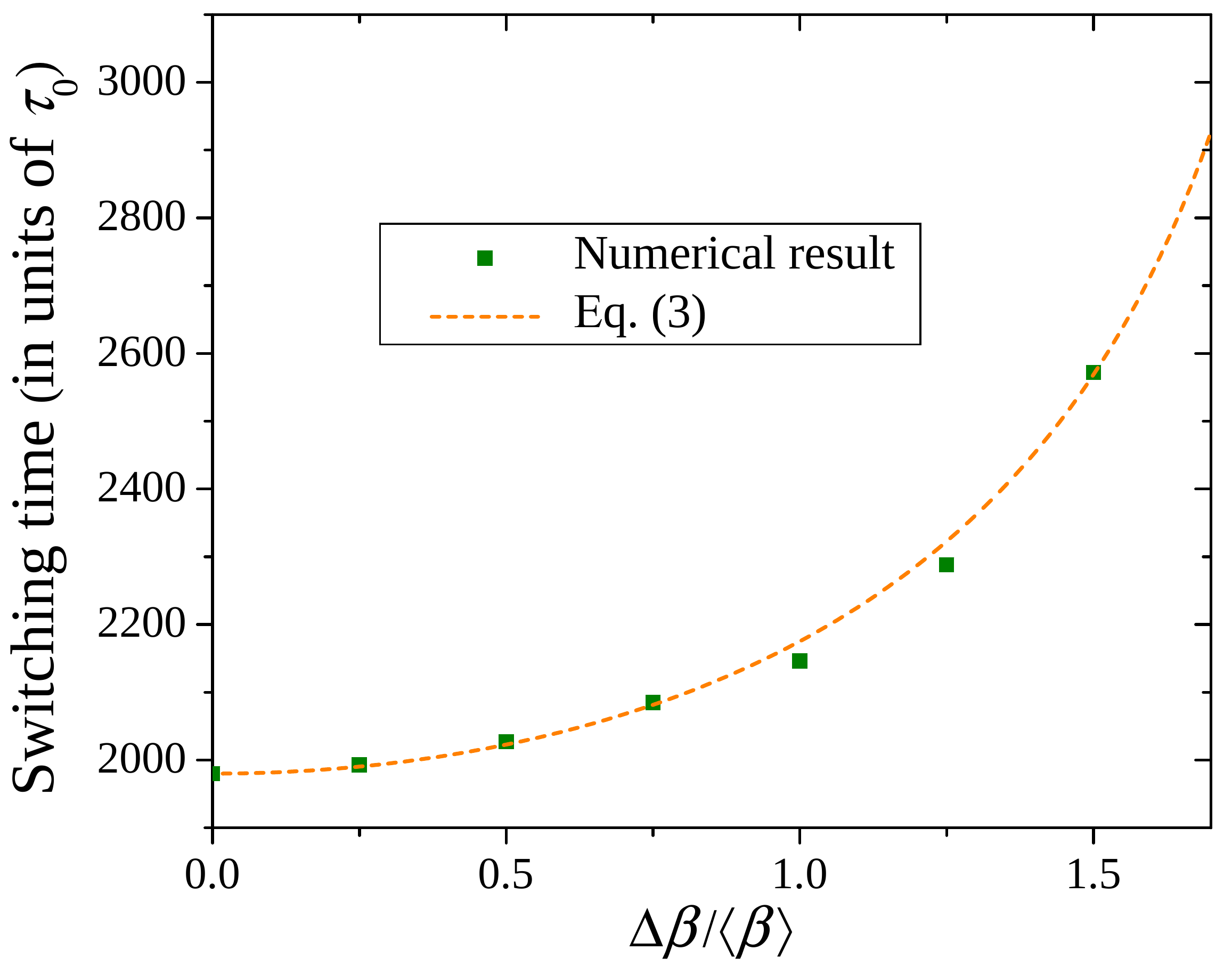}
\caption{\label{fig3} (Color online) Network switching time as a function the distribution width $\Delta\beta$. The switching times have been calculated for a network of $N=1000$ memristive systems with a flat random distribution of switching constants subjected to $V=1.05NV_t$. The dashed curve is plotted assuming that the switching time is proportional to $\langle 1/\beta \rangle$ given by Eq. (\ref{eq:harmonic_av}).}
\end{center}
\end{figure}

Fig. \ref{fig2}(b) also demonstrates that the total switching time for the entire network is longer than the switching time defined by the average value of $\beta_i$-s, which is  $\langle \beta\rangle$. This general type of behavior has been verified for many different realizations of memristive systems. Fig. \ref{fig3} shows some additional data points (extracted from numerical simulations) demonstrating a monotonic increase in the network switching time with distribution width.  In fact, according to our analytical theory presented below, the network switching time is actually proportional to $\langle 1/\beta \rangle$ instead of $1/\langle \beta \rangle$ (See Eq. \ref{eq:switch t}). For a flat random distribution of $\beta_i$ in the range $[\langle\beta\rangle-\Delta \beta/2,\;\;\langle\beta\rangle+\Delta \beta/2]$ and $N\gg1$
\begin{equation}
 \langle 1/\beta \rangle=\frac{1}{\Delta \beta}\ln\frac{\langle\beta\rangle+\frac{\Delta \beta}{2}}{\langle\beta\rangle-\frac{\Delta \beta}{2}} \label{eq:harmonic_av}.
\end{equation}
The dashed curve in Fig. \ref{fig3} shows the perfect agreement of Eq. (\ref{eq:harmonic_av}) with our numerical results.

If the distribution of $\beta_i$-s is not flat then one can show that for {\it any} distribution of
$\beta_i$ the difference $\langle 1/\beta\rangle-1/\langle\beta\rangle\geq 0$.
Additionally, this difference normally grows with the distribution width.
For example, if all odd central momenta are negative or equal to zero (as,
for example, for the Gaussian distribution), then this difference cannot be less than
$(\langle\beta^2\rangle-\langle\beta\rangle^2)/\langle\beta\rangle^3$. Thus, our observation of the switching time increase is valid on average for any distribution of $\beta_i$.

{\it Analytical Model.} -- If the initial memristances are the same and the applied voltage exceeds the combined threshold voltage $NV_t$, then one can realize that the voltage fall across any memristive system exceeds its threshold voltage at any time. Moreover, in the case of a distribution of initial memristances, the same is true either from $t=0$ or after an initial equilibration period. Therefore, in the region of parameters of interest, Eq. (\ref{eq:model2}) can be generally written as
\begin{equation}
\dot{R}_i(t)=-\beta_i\left[ V_i(t)-V_t\right], \label{eq:1}
\end{equation}
where $i=1,...,N$, $V_i(t)=VR_i(t)/R(t)$, and $R(t)=\sum\limits_{i=1}^{N}R_i(t)$ is the total memristance. The minus sign in Eq. (\ref{eq:1}) takes into account the selected connection polarity of memristive systems in the network (such that their memristances decrease at positive applied voltages).

\begin{figure}[tb]
\begin{center}
\includegraphics[width=.95\columnwidth]{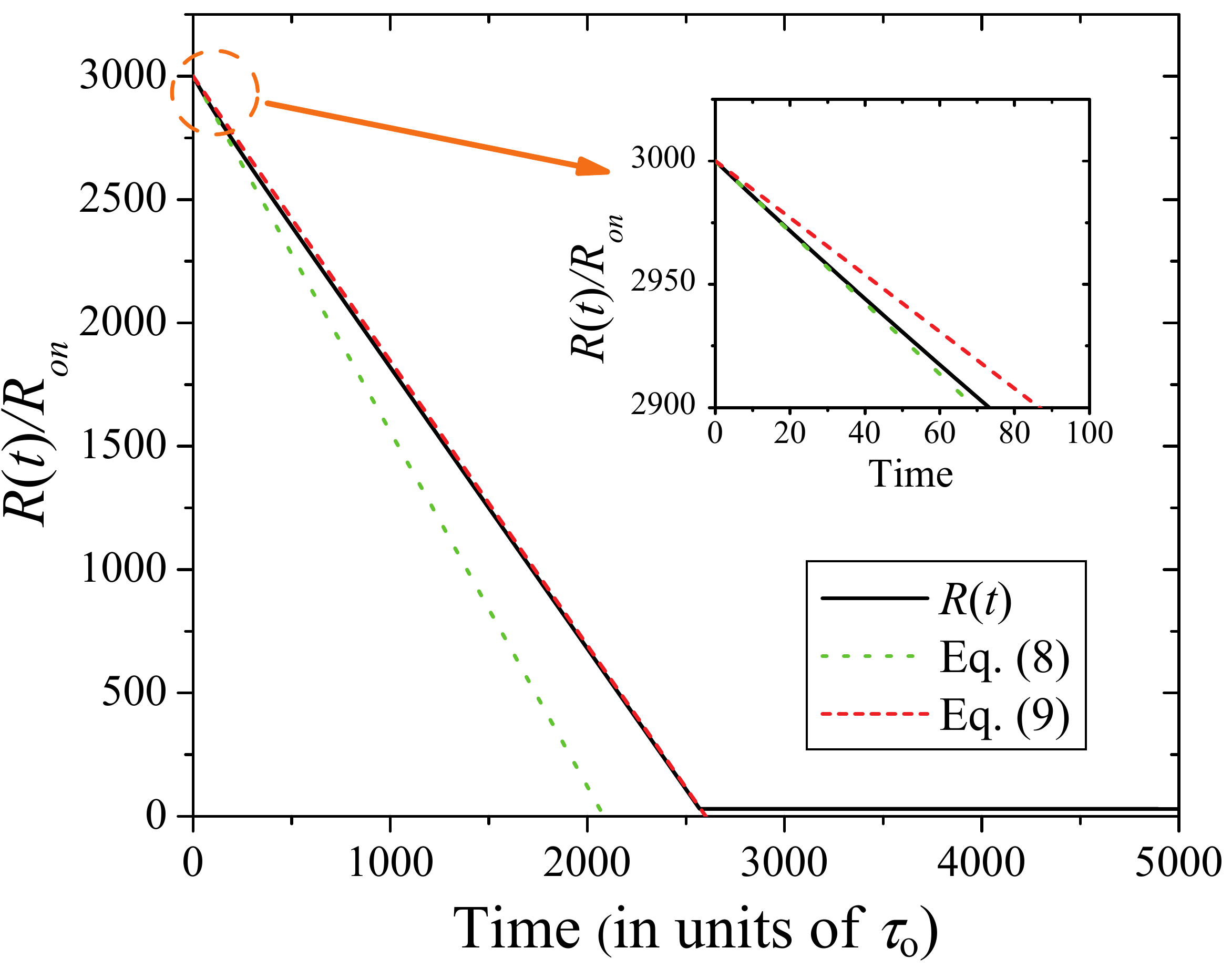}
\caption{\label{fig4} (Color online) Comparison of asymptotic expressions (Eqs. (\ref{eq:small t}) and (\ref{eq:5})) for the time-dependence of the total memristance $R(t)$ and numerical (exact) solution. The numerical solution (solid black curve) has been obtained  for the same realization of memristive systems, model and simulation parameters as in Fig. \ref{fig2}. Eq. (\ref{eq:small t}) curve is plotted in the linear approximation.}
\end{center}
\end{figure}

Let us search for the solution of Eq. (\ref{eq:1}) in the form
\begin{equation}
R_i(t)=C_i(t)e^{-\beta_i\int\limits_0^t\frac{V}{R(\tau)}\textnormal{d}\tau}, \label{eq:2}
\end{equation}
where $C_i(t)$ is a time-dependent function and the integral in the exponent is actually the charge ($\int\limits_0^tV/R^{-1}(\tau)\textnormal{d}\tau=q(t)$) flown through the network by the time $t$. Substituting Eq. (\ref{eq:2}) into Eq. (\ref{eq:1}), one can find the following expression for $R_i(t)$:
\begin{equation}
R_i(t)=R_i(0)e^{-\beta_iq(t)}+\beta_iV_te^{-\beta_iq(t)}\int\limits_0^{t}e^{\beta_iq(t')}\textnormal{d}t'. \label{eq:3}
\end{equation}
Taking into account that $R_i(0)=R_{off}$, the sum of Eqs. (\ref{eq:3}) yields
\begin{equation}
R(t)=\sum\limits_{i=1}^N e^{-\beta_iq(t)}
\left[R_{off} +\beta_iV_t \int\limits_0^{t}e^{\beta_iq(t')}\textnormal{d}t'\right]. \label{eq:4}
\end{equation}
As $q(t)$ can be expressed through $R(t)$ (see the definition of $q(t)$ below Eq. (\ref{eq:2})), Eq. (\ref{eq:4}) can be considered as the nonlinear integral  equation for $R(t)$.

While it is difficult to find the exact solution $R(t)$  from Eq. (\ref{eq:4}), this equation can be effectively used to derive the asymptotic behavior of $R(t)$ in the most important limiting cases. In particular, in the short time limit, one can expand $R(t)=R_{0}(1-at+b t^2+O(t^{3}))$
using unknown constants $a$ and $b$, and get $q(t)= V(t+at^{2}/2+O(t^{3}))/R_{0}$.  Using Eq. (\ref{eq:4}) one can find
\begin{equation}
R(t)=NR_{off}-\langle\beta\rangle\delta V t+ \frac{D\beta V\delta V t^2}{2NR_{off}}+O(t^{3}),
~t\rightarrow 0,
\label{eq:small t}
\end{equation}
where $\delta V=V-NV_t$ is the voltage excess above the combined threshold, and $D\beta=\langle\beta^2\rangle-\langle\beta\rangle^2$ is the dispersion
of switching constants $\beta_i$. While the first and the second terms in Eq. (\ref{eq:small t}) correspond to the averaged effect of  evolution of individual memristive systems, the third term, being proportional to the dispersion, is always positive and describes effect of collective evolution of memristive network. Note that the expression (\ref{eq:small t}) is valid only when
the second and  third  terms are small compared to the first one.

A different asymptotic expression can be found in the long time limit, namely, when $\beta_iq(t)\gg 1$. This limit also implies the optimal synchronization condition $\delta V\ll V$ as demonstrated below. When $\beta_iq(t)\gg 1$, the main contribution to the right-hand side of Eq. (\ref{eq:4}) comes from the upper limit of the integral with respect to $t'$. Using this observation one can derive the following main term of the long time asymptotic
\begin{equation}
R(t)=(NR_{off}- \beta_H \delta V t)(1+O(\delta V/V)),
~\delta V\rightarrow +0,
\label{eq:5}
\end{equation}
where  $\beta_H=\langle 1/\beta\rangle^{-1}$.

To specify the applicability conditions of Eq. (\ref{eq:5}), one can calculate $q(t)$ using Eq. (\ref{eq:5}). Then the condition
 $\beta_iq(t)\gg 1$
can be presented as
\begin{equation}
\frac{\beta_i V}{\beta_H \delta V}\ln\left(\frac{NR_{off}}{R(t)}\right)\gg 1.
\label{eq:large t}
\end{equation}
Eq. (\ref{eq:large t}) can be sub-divided into the optimal synchronization condition $\delta V\ll V$ (also observed in our numerical studies) and the condition of long times such that the total resistance $R(t)$ is much less than its initial value $NR_{off}$. The total switching time $T$ for the network
can be easily computed  substituting $R(T)=NR_{on}$ in Eq. (\ref{eq:5}). This gives
\begin{equation}
T=\frac{N(R_{off}-R_{on})}{ V-NV_t}\left\langle\frac{1}{\beta}\right\rangle,
~\delta V\ll V.
\label{eq:switch t}
\end{equation}
Furthermore, in the typical situations when $R_{off}\gg R_{on}$, $R_{on}$ in the nominator of Eq. (\ref{eq:switch t}) can be omitted.

Fig. \ref{fig4} shows a comparison of the numerically obtained solution with the asymptotic expressions given by Eqs. (\ref{eq:small t}) and (\ref{eq:5}). Clearly, the asymptotic expressions are in the excellent agreement with the numerical solution for $R(t)$.

It is interesting to note that Eq. (\ref{eq:5})
also delivers a good approximation for all times  when $\delta V\ll V$.
This allows to find the approximate expression for  individual resistances
$R_i(t)$ for all moments
of time, which reproduces exactly the asymptotic behavior  (\ref{eq:5}) for
long times and
the first two terms of (\ref{eq:small t}) for short times for the total resistance
$R(t)$.
Thus, an approximated expression for $R_i(t)$ can be obtained substituting Eq. (\ref{eq:5}) into Eq. (\ref{eq:3}). Assuming that $\delta V\ll V$, which is the optimal condition for the synchronization,
one can obtain
\begin{eqnarray}
R_i(t)=R_{off}-\frac{\beta_H \delta V}{N}t-R_{off}\left( 1-\frac{\beta_H}{\beta_i}\right)\frac{\delta V}{V_t N} \nonumber \\
+R_{off}\left( 1-\frac{\beta_H}{\beta_i}\right)\frac{\delta V}{V_t N} e^{-\frac{\beta_i V}{R_{off} N}t}. \;\;\; \label{eq:6}
\end{eqnarray}
The exponential (last) term in the second line of Eq. (\ref{eq:6}) decays on a short time scale.
Clearly,
the ratio of this short time scale to the total switching time
$T$, Eq. (\ref{eq:switch t}), is  $\delta V/V \ll 1$.

Moreover, it is easy to notice that the first two terms in the right-hand
side of  Eq. (\ref{eq:6}) are dominant at long times.
These terms are the same for all memristive systems (they do not depend on the system index $i$) and therefore the memristances of the memristive systems in the network are nearly the same.
This observation confirms our numerical results (see, Fig. \ref{fig2}(b)).

{\it Conclusion.}-- In conclusion, we have discovered an interesting synchronization effect taking place in one-dimensional memristive networks with elements characterized by a distribution of switching constants. When the switching occurs from the high to low resistance state, the systems with larger switching constants slow down their switching as the voltage falls across these systems decrease faster compared to voltages across the systems with lower switching constants. As a result, the switching of all memristive systems occurs coherently with the same effective rate regardless the specific switching constants of individual systems. This simple picture explains the mechanism of the synchronization effect that is most pronounced when the applied voltage slightly exceeds the combined threshold voltage of memristive systems. We have also demonstrated that the network switching time is independent on the number of memristive systems (for an appropriately scaled applied voltage) and is defined by the harmonic mean of switching constants.

This work has been supported by the NSF grant No. ECCS-1202383.

\bibliography{memcapacitor}

\end{document}